\documentstyle[aps,prl,twocolumn,epsf]{revtex}

\begin{document} 

\title{Influence of the temperature on the depinning transition of
  driven interfaces}
\author{U.~Nowak \cite{uliEmail} and K.~D.~Usadel \cite{usadelEmail}}
\address{Theoretische Tieftemperaturphysik,
  Gerhard-Mercator-Universit\"{a}t-Duisburg, 47048 Duisburg/ Germany\\}

\date{\today}
\maketitle


\begin{abstract}
  We study the dynamics of a driven interface in a two-dimensional
  random-field Ising model close to the depinning transition at small
  but finite temperatures $T$ using Glauber dynamics. A square lattice
  is considered with an interface initially in (11)-direction. The
  drift velocity $v$ is analyzed using finite size scaling at $T=0$
  and additionally finite temperature scaling close to the depinning
  transition. In both cases a perfect data collapse is obtained from
  which we deduce $\beta \approx 1/3$ for the exponent which
  determines the dependence of $v$ on the driving field, $\nu \approx
  1$ for the exponent of the correlation length and $\delta \approx 5$
  for the exponent which determines the dependence of $v$ on $T$.\\
  
\end{abstract}


The motion of a driven interface in a random medium has attracted a
large amount of interest recently because it is a challenging
theoretical problem and because it occurs in many different areas of
physics like fluid invasion of porous media \cite{JiRo1}, depinning of
charge density waves \cite{Fisher} or field driven motion of a domain
wall in a ferromagnet \cite{ApBru}, to name just a few. In these cases
the disorder is time independent or quenched. For zero temperature
there is a well defined critical driving force $F_c$ above which the
interface moves while the interface gets trapped in metastable
positions below it. This scenario has been observed in a variety of
different works in the past \cite{review}.

At finite temperature the depinning transition is smeared out and the
interface can move also below $F_c$. For very small driving forces a
creep motion is expected which is governed by thermal activation while
for driving forces close to the critical one a scaling behavior of the
interface velocity $v$ is expected to occur for small enough
temperatures (see \cite{Fisher} where the influence of the temperature
on the velocity of sliding charge-density waves is discussed).

The dynamics and the morphology of interfaces in systems with quenched
disorder have been investigated in the past within a variety of
different models.  Most studies focus on equations of motion for the
interface itself. Famous approaches are the Edwards-Wilkinson (EW)
equation \cite{ew} or the Kardar-Parisi-Zhang (KPZ) equation
\cite{kpz} which have been studied extensively both with annealed and
quenched disorder.  A simple example for a disordered medium in which
the motion of an interface can be studied is the random-field Ising
model.  In this case the interface is a domain wall separating regions
of up and down spins.  For this system it were Bruinsma and Aeppli
\cite{ApBru} who argued with the assumption that the interface can be
treated as an elastic membrane that its equation of motion at the
depinning transition can be described by the EW equation with quenched
disorder.  Their arguments are plausible but far from rigorous.
Therefore, it is also of great interest to tackle the full problem,
i.~e.~to study the dynamics associated with the Hamiltonian of the
random-field Ising model prepared initially with an interface. Under
the influence of external forces the domain wall may begin to move in
a way which depends on the strength of the disorder. A still unsolved
problem is the scaling behavior of the velocity of the domain wall for
small but finite temperature close to the critical driving field.
This, of course, is an important issue also experimentally
\cite{Nowak}.

It is the purpose of this letter to elucidate this critical behavior
of a domain wall in a two-dimensional random-field Ising model close
to the pinning transition. Finite size scaling is used for the
analysis of the zero temperature behavior resulting in precise values
of the exponents. For the first time we determine the critical
exponent $\delta$ associated with the finite temperature smearing of
the transition which to the best of our knowledge is unknown at
present.

For a random-field Ising system in previous work \cite{JiRo2} a
cubic structure was considered with an interface between up and down
spins initially parallel to one of the cubic axis of the lattice. An
applied magnetic field $H$ favoring the up spins energetically leaves
the down spins in a metastable state. Under the influence of some
dynamical rules the area occupied by metastable states shrinks, i.~e.
the interface starts to move. For the dynamics at zero temperature
simple relaxation dynamics was assumed, i.~e. a spin is flipped only
if its energy is lowered. Additionally it was assumed that only spins
at the interface are allowed to flip in order to avoid spontaneous
domain growth in the metastable state. The critical field was
approached from below, i.~e.~the driving field was increased in small
steps until the pattern of flipped spins after an increase of the
field spans the system. The field value for this to happen is the
critical field $H_c$ .

The geometry used in these investigations has the disadvantage that
even without impurities the interface is pinned for up to quite large
fields. For space dimension $d=2$, for instance, the driving field has
to be larger than $2J$ for the interface to move where $J$ denotes the
nearest neighbor exchange interaction. But if one spin of the
interface is flipped all neighbors in the same line will flip in the
following updates. Thus for zero or very small disorder the movement
of the interface consists of a series of complete lines of spins which
flip resulting in faceted growth. In this case pinning of the
interface is not due to disorder but to the fact that any spin in the
interface is locked by its four neighbors.  Faceted growth can occur
in any dimension for weak and bounded disorder but it depends on the
structure of the lattice \cite{KoJiRo}.

By changing the geometry, i.~e.~considering a (11)-interface instead,
it is possible to have an interface which without disorder moves for
arbitrarily small driving fields. The dynamics of the interface is
then well defined for any strength of the disorder. However the most
important observation is that now the conventional $T=0$ dynamics in
which a spin flips if this lowers the energy of the system can be
generalized to finite temperatures. Since for weak disorder only small
driving fields are needed at low temperatures conventional Glauber
dynamics can be used without running into the problem of spontaneous
domain growth in the metastable phase. For small, bounded disorder
there occurs a separation of time scales in the sense that for the
system considered within the necessary simulation time only interface
motion is observed. The time scale for which the system runs into
thermal equilibrium - which Glauber dynamics necessarily does of
course - is many orders of magnitude larger, for more details see
below. This clear separation of time scales makes a numerical study of
temperature effects close to the pinning transition feasible.

The random-field Ising model we consider is defined by the Hamiltonian
\begin{equation}
{\cal H} = -J\sum_{<ij>} \sigma_i \sigma_j - H \sum_i \sigma_i  - \sum_i H_i
\sigma_i,
\end{equation}
where $\sigma_i=\pm 1$ are Ising spins on a two dimensional lattice.
The first sum describes the ferromagnetic nearest neighbor interaction
($J > 0$). The random fields $H_i$ are taken from a distribution which
is constant within an interval $[-\Delta:\Delta]$ and hence are
bounded. $H$ is the homogenous driving field.

\begin{figure}
  \begin{center}
    \epsfxsize=6cm
    \hspace*{7mm}
    \epsffile{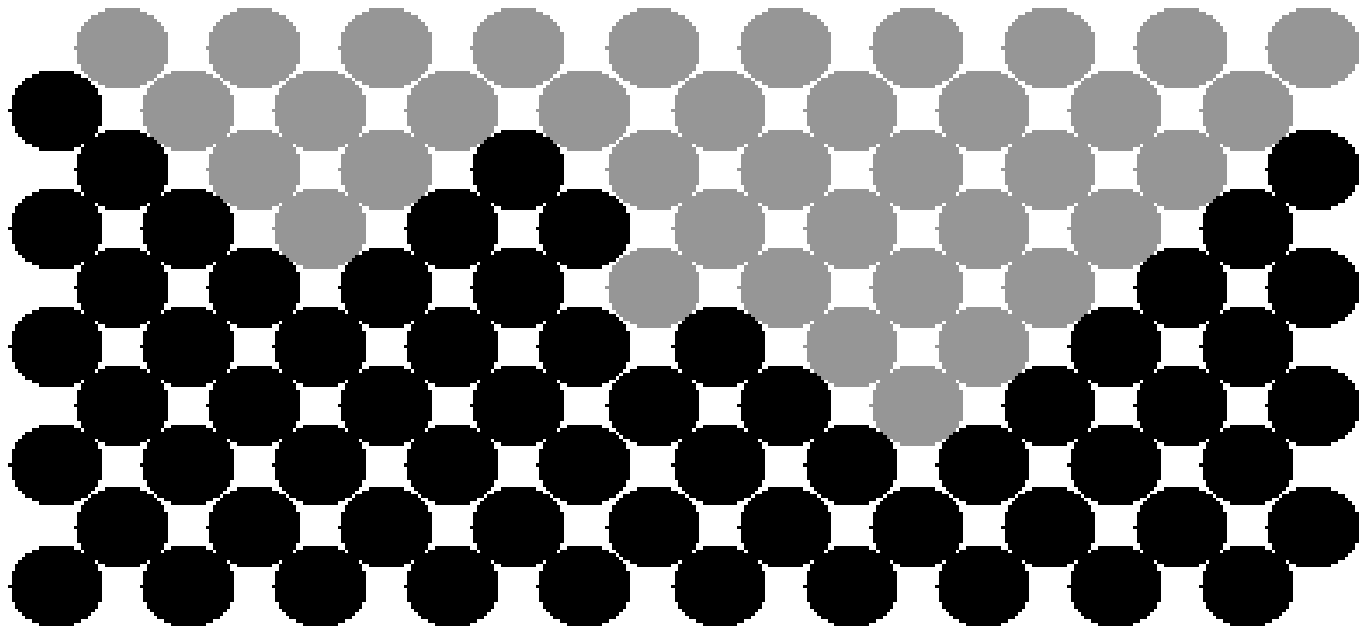}

    \hspace*{7mm}
    \epsfxsize=6cm
    \epsffile{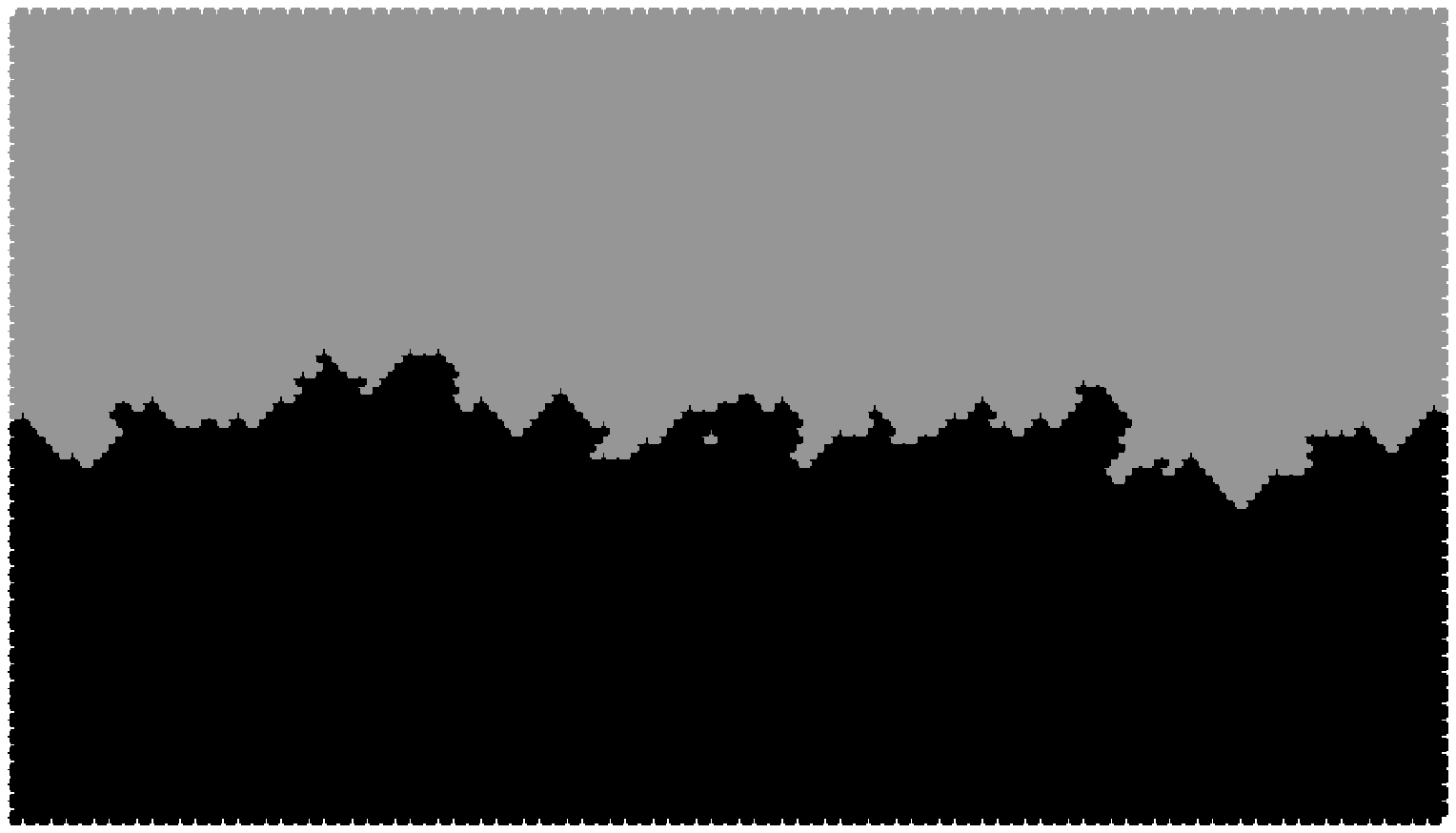}
  \end{center}
  \caption{a) (upper picture) Schematic picture of a domain wall in
    the simulated lattice configuration. b) (lower picture) Domain
    wall of a simulated system of size $100 \times 100$ at the
    depinning field ($H_c = 1.29J$) for finite temperature ($T=0.1J$).}
  \label{f:wall}
\end{figure}

The lattice we consider is shown in Fig.~\ref{f:wall}a). It has the
structure of a square lattice but it is rotated by an angle of
$\pi/4$. Hence, the boundaries and the initially flat interface of the
system are in (11)-direction. We use periodic boundary conditions
parallel to the interface. Therefore all spins at the initial
interface have exactly two of its four nearest-neighbor bonds broken,
i.~e. it does not cost any exchange energy to flip a spin at the
interface. Hence, without disorder there is no pinning of the
interface.

In the direction perpendicular to the interface we use fixed boundary
conditions. Spins in the lowest line are always up (in direction of
the driving field) and in the last line are always down.  At the
beginning of the simulation all spins except of the lowest line are
down and, hence, the interface is above the first line.  Later, during
the simulation the interface moves up due to the driving field.

For finite temperature we perform Monte Carlo simulations where the
spins are updated randomly using a heat-bath algorithm.  For zero
temperature this algorithm naturally crosses over to an algorithm
where spins are only flipped when its energy decreases when flipped.

Fig.~\ref{f:wall}b) shows a system of size $100 \times 100$ spins
during the simulation. The strength of the random field is
$\Delta=1.5J$ and we will use this value throughout the whole paper.
The driving field is $H = 1.29J$. We will show later that this value
is the critical depinning field for $\Delta = 1.5J$ at zero
temperature. The interface seen is not pinned due to a finite
temperature of $T=0.1J$ used in this simulation. No thermal
fluctuation or domains appear in the unstable phase (upper part).
This is important for the following reasons. In equilibrium the two
dimensional random-field system has no long range ordered phase
\cite{Aizen}.  Since Monte Carlo simulation in principal leads to
equilibrium properties one could expect that the system splits into
domains spontaneously so that the concept of a single domain wall
within the system is no longer useful. But this is not the case due to
a separation of time scales mentioned above. Within the single spin
flip method the growth of a domain must start with flipping one spin.
The minimum energy needed for this process is $\Delta E =
2(4J-H-\Delta)$ which is $3.42J$ here for the critical driving field.
The corresponding flipping probability within a Monte Carlo simulation
is $\exp(-\Delta E/T) \approx 7 \times 10^{-14}$ for $T=0.1J$.
Therefore, a thermal fluctuation within the bulk is far from being
possible within our simulation time which is less than $10^5$ Monte
Carlo steps per spin (MCS) -- it happens on much larger time scales.
Hence, as long as we restrict ourselves to low enough temperatures we
can perform simulations without the possibility of spontaneous domain
growth in the metastable phase.

The velocity of the domain wall we define as the time derivative of
the magnetization in a steady state. Note that at the beginning of the
simulation the wall is flat and consequently the system is not in a
steady state. In the steady state the magnetization grows linearly
with time with fluctuations from sample to sample, of course. The
velocity is obtained as averaged slope in this linear region. We
average the velocity over many systems (10-160, depending on system
size and how close the value of the driving field is to the critical
one) and perform simulations for different system sizes ($24 \times
400 \ldots 400 \times 400$) in order to investigate finite size
effects systematically.

\begin{figure}
  \begin{center}
    \epsfysize=4.5cm
    \epsffile{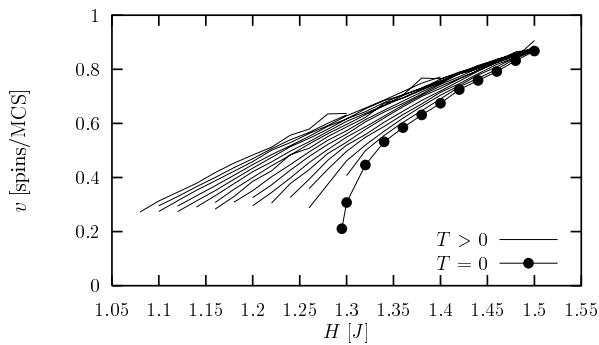}
  \end{center}
  \caption{$v$ versus $H$ for $T=0$ and for finite temperatures ranging
    from 0.01$J$ to 0.16$J$. The system size is $400 \times 400$.}
  \label{f:vvh_t}
\end{figure}

Our results for the domain wall velocity $v$ versus driving field are
shown in Fig.~\ref{f:vvh_t} for different finite temperatures and
$T=0$. During this simulation we consider for each temperature only
driving fields which are large enough so that the wall crosses the
system within roughly 
\begin{figure}
  \begin{center}
    \epsfysize=4.2cm
    \epsffile{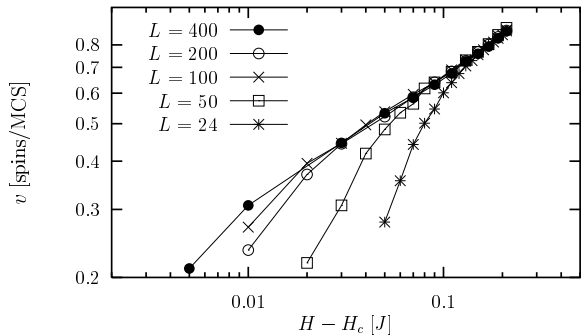}
  \end{center}
  \caption{$v$ versus $H-H_c$ with $H_c = 1.29J$ determined from
    finite size scaling (see Fig.~4). $T=0$.}              
  \label{f:vvh}
\end{figure}
2000MCS. Otherwise we interrupt the simulation.
As Fig.~\ref{f:vvh_t} suggests we get a nice depinning transition at
$H_c \approx 1.29J$ which is smeared out for finite temperatures.

The wall velocity $v$ depends on system size as shown in
Fig.~\ref{f:vvh} for $T=0$. Finite size effects are common at
equilibrium phase transitions and there it is known that the most
reliable values for the critical exponents are deduced from a finite
size scaling analysis. We found that finite size scaling also works in
the present case.  For zero temperature the wall can be pinned and
within these simulations we set the velocity of a single system to
zero if the wall did not cross the system within roughly 20000 MCS and
we interrupt the simulation when more then one half of the simulated
systems have velocity zero.  In contrast to the situation at ordinary
phase transitions here the effect of finite size is reversed: the
smaller the systems are the stronger vanishes the order parameter. We
use the finite-size scaling ansatz

\begin{equation}
  v(L,H) = L^{-\beta/\nu} \tilde{v}\left((H-H_c)L^{1/\nu}\right),
\end{equation}

where for the scaling function $\tilde{v}$ we demand $\tilde{v}(x)
\sim x^{\beta}$ for $x \gg 1$ and --- in contrast to ordinary phase
transitions --- $ \tilde{v}(x) = 0$ for $x \rightarrow 0$.

\begin{figure}
  \begin{center}
    \epsfysize=4.5cm
    \epsffile{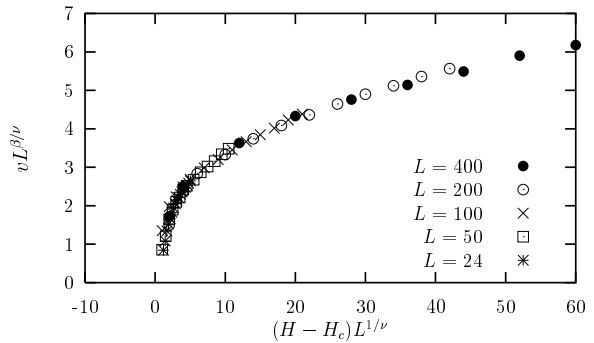}
  \end{center}
  \caption{Finite size scaling plot of Fig.~\ref{f:vvh}.}
  \label{f:fisi}
\end{figure}

Fig.~\ref{f:fisi} demonstrates that the finite size scaling ansatz
above works well. Note that the scaling function has the unusual
property mentioned above, i.~e. it goes to zero for $H \rightarrow
H_c$.  From the finite size scaling we get $H_c = 1.290J \pm 0.003J$,
$\beta = 0.35 \pm 0.04$, and $\nu = 1.0 \pm 0.05$.  Alternatively,
these results for $\beta$ and $H_c$ can also be obtained directly from
Fig.~\ref{f:vvh} by fitting those data which do not show finite size
effects to a power law $v \sim (H-H_C)^{\beta}$. The corresponding
line is also shown in Fig.~\ref{f:vvh}. Note also, that this value for
$\beta$ is in agreement with the earlier result in \cite{Nolle} within
the error bars.

\begin{figure}
  \begin{center}
    \epsfysize=4.cm \epsffile{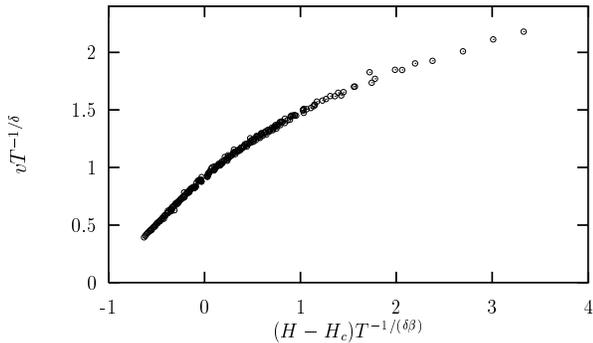}
  \end{center}
  \caption{Scaling plot corresponding to Eq.~\ref{e:tscal} for
    the data of Fig.~\ref{f:vvh_t}.}
  \label{f:vt_skal}
\end{figure}

In order to analyse the finite temperature effects we first note that
for the largest system size we investigated ($400 \times 400$) there
are no finite size effects for velocities $v$ well above 0.2 as
suggested by Fig.~\ref{f:vvh}. For $L=400$ all data points except the
one close to 0.2 are on a straight line, i.e. increase as a power law
without $L$-corrections. Hence, it is safe to neglect size effects for
the finite temperature data in Fig.~\ref{f:vvh_t}. For the smearing of
the transition by temperature we expect a scaling behavior
\cite{Fisher}
\begin{equation}
  v(H,T) = T^{1/\delta} \tilde{v}\left((H-H_c)T^{-1/(\beta\delta)}\right).
\label{e:tscal}
\end{equation}
A corresponding scaling plot of our data is shown in
Fig.~\ref{f:vt_skal}.

\vspace{5mm}

We obtain a perfect data collapse with the following parameters: $H_c
= 1.290J \pm 0.005J$, $\beta = 0.33 \pm 0.02$, and $\delta = 5.0 \pm
0.3$. To the best of our knowledge this is the first work in which the
exponent $\delta$ is determined for the system considered. Note that
the perfect data collapse observed leads to exponents with only very
small statistical errors.

Finite size scaling and finite temperature scaling both give
independent values for the critical field $H_c$ and for the exponent
$\beta$, respectively. It is rather satisfying that in both cases the
numerical values agree within the error bars.

As was mentioned in the introduction there are arguments in favor of
the conjecture that the motion of a domain wall in a random-field
system can be described by the EW equation with quenched
disorder\cite{ApBru} at the depinning transition. For this equation
the scaling relation $ \nu (2 - \zeta) = 1$ has been derived
\cite{Nattermann} by a functional renormalization group scheme in
$4-\epsilon$-dimensions. This scaling relation has been claimed to be
valid to all orders of $\epsilon$ \cite{Narayan}.  If we adopt this
view we are therefore able to determine the roughness exponent of the
random-field system without investigating the morphology of the
interface. The value we obtain, $\zeta \approx 1$, agrees with the
value obtained from an $\epsilon$-expansion which is also claimed to
be exact in all orders of $\epsilon$ \cite{Nattermann,Narayan}. Values
of this exponent obtained numerically by integrating the EW equation
or an automaton version of it \cite{Lesch} scatter between $0.7$ and
$1.25$ \cite{expo,Lesch}.  From the scaling relation $\nu (z - \zeta)
= \beta$ we get $z \approx 4/3$ for the dynamic exponent which - as
well as our result for $\beta$ - is also in agreement with the results
of Nattermann et al. \cite{Nattermann}. Interestingly, our value for
the temperature exponent $\delta$ fulfills the scaling relation
$\delta = 2 + 1/\beta$ which has been derived by Tang and Stepanow
\cite{Tang} by an extension of the functional renormalization group
scheme mentioned above to finite temperatures. All our findings
support that the motion of a domain wall in a random-field system can
be described by the EW equation. However, for an interface moving with
a finite velocity also KPZ-like non-linearities could become relevant
but from our numerical data we cannot extract any conclusions
concerning a corresponding crossover of the values of exponents or
scaling laws.

To conclude, we investigated the influence of finite temperatures on
the depinning transition of a driven [11]-interface by a Monte Carlo
simulation of the two dimensional random-field Ising model.  We used
bounded disorder and low temperatures in which case a clear separation
of time scales occurs in the sense that no spontaneous growth of
domains in the metastable phase appears. We derived the order
parameter exponent $\beta \approx 1/3$ as well as the exponent $\nu
\approx 1$ of the correlation length using finite size scaling.  The
corresponding dynamic exponent $z \approx 4/3$ and roughness exponent
$\zeta \approx 1$ are determined via scaling relations. The exponent
$\delta$ describing the influence of the temperature on the depinning
of the domain wall which was unknown before was determined to be
$\delta \approx 5$.

{\bf Acknowledgments:} The work was supported by the Deutsche
Forschungsgemeinschaft through Sonderforschungsbereich 166 and through
the Graduiertenkolleg "Heterogene Systeme". We also thank
T.~Nattermann for pointing us to Ref.~\cite{Tang}, and S.~Stepanow
and L.~H.~Tang for providing their preprint \cite{Tang} prior to
publishing.

\end{document}